\documentclass[aps,twocolumn,prb,10pt]{revtex4-1}
\setcitestyle{super,open={},close={}}
\usepackage{amsmath}
\usepackage{amssymb}
\usepackage[pdftex]{graphicx}
\usepackage[usenames]{color}
\usepackage{mdframed}

\definecolor{orange}{rgb}{1,0.5,0}

\newcommand{\R}[1] {{\color{black} #1}}

\begin{document}

\title {Using Azobenzene Photocontrol to Set Proteins in Motion}
\author{Olga Bozovic, Brankica Jankovic, *Peter Hamm  \\\textit{Department of Chemistry, University of
		Zurich, Zurich, Switzerland}\\  $^*$peter.hamm@chem.uzh.ch}
	
\begin{abstract}
Controlling the activity of proteins with azobenzene photoswitches is a potent tool for manipulating their biological function. With the help of light, one can change e.g. binding affinities, control allostery or temper with complex biological processes. Additionally, due to their intrinsically fast photoisomerisation, azobenzene photoswitches can serve as triggers to initiate out-of-equilibrium processes. Such switching of the activity, therefore, initiates a cascade of conformational events, which can only be accessed with time-resolved methods. In this Review, we will show how combining the potency of azobenzene photoswitching with transient spectroscopic techniques helps to disclose the order of events and provide an experimental observation of biomolecular interactions in real-time. This will ultimately help us to understand how proteins accommodate, adapt and readjust their structure to answer an incoming signal and it will complete our knowledge of the dynamical character of proteins.\\

%\centering{TOC Graphics:}
%
%\centering\includegraphics[width=0.4\textwidth]{FigTOC.pdf}

\end{abstract}

\maketitle
\section{Introduction}

Proteins have a dynamical nature that governs their biological function, and understanding the underlying processes ultimately allows for a fine-tuning of the affinity, controlling allostery, or regulating the activity of target proteins. When discussing the dynamical nature of proteins, it is essential to distinguish between equilibrium and non-equilibrium dynamics.\cite{Henzler-Wildman2007} Equilibrium dynamics of proteins lie in the fact that native proteins, even when fully folded, are not rigid entities.\cite{Dyson1998,Bernado2010} The equilibrium dynamics of proteins is reflected by an ensemble of conformational sub-states and the continuous fluctuation between them. The prototype example illustrating the importance of protein fluctuations is myoglobin. The heme is deeply buried inside the protein and oxygen would not be able to access it if the structure of myoglobin would be the static X-ray representation.\cite{elber90} Studying the equilibrium dynamics of proteins is mostly the realm of NMR spectroscopy \R{(see Table~\ref{tab1})}.\cite{Palmer2004a,Mittermaier2006,Kay2005}

Non-equilibrium dynamics, on the other hand, describe how proteins adapt and respond to an incoming event. This event can be the interaction with a signalling molecule or another protein, or the response to an allosteric signal.\cite{Olsson2006,Benkovic2003} In order to experimentally follow the non-equilibrium adaptation of proteins to a signal, one needs an instantaneous perturbation that initiates a response of the system while following a time-dependent observable. The non-equilibrium dynamics portray how a system behaves in real-time, provides us with a ``molecular movie'', and broadens our in-equilibrium picture of a molecular mechanism. The static representation of biomolecular recognition, protein allostery or enzymatic activity, as for instance the ``lock and key'' mechanism, have long been known to be inadequate, given that they do not represent the natural dynamical fluidity of proteins and their interacting partners.\cite{csermely2010} Even for an event as ``simple'' as small ligand binding to a protein, many complex, transient and sometimes subtle redistributions of interactions and conformational changes do occur.

Complex biological processes, such as enzymatic and metabolic cascades within diverse signaling pathways, involve the communication between many players and as such require many underlying layers of control. Consequently, there are several mechanisms for biological regulation of protein’s activity, where the targets for regulation can be an orthosteric or an allosteric site. Orthosteric regulation represents a direct regulation of biomolecular recognition, i.e., of protein-ligand interactions at the active/binding site. It is manifested by the competitive binding of different ligands or by e.g., a reversible phosphorylation that alters the binding affinity. Allosteric regulation, on the other hand, implies binding of a signaling molecule on a site different from an active site. The communication between two non-overlapping sites regulates the activity of the protein by an``allosteric signal'', the nature of which is hotly debated.\cite{swain2006,Tsai2008,smock2009,Tsai2014,Hilser2012} Implementation of time-resolved spectroscopic techniques is of paramount importance for a full dynamical description of such biological processes.

\begin{table*}
\centering
\caption{\R{Comparison of various structure sensitive spectroscopic methods to study the dynamics of proteins.}}
\label{tab1}
\begin{tabular}{l | l |l| l }
                                           & NMR & time-resolved IR & time-resolved XRD \\\hline
Sample                                     &solution ($\sim100~\mu$M) & solution ($\sim1~m$M)  & crystal     \\
Overall structure resolution power         &high	        &low	                          &high	\\
Local structure resolution power           &high	        &very high\footnotemark[1]        &high   \\
Time-range for equilibrium dynamics        &$\sim$100~fs-1~s      &$\sim$100~fs-10~ps     &--  \\
Time-range for non-equilibrium dynamics    &$\gtrsim$100~ms        &$\gtrsim$1~ps &$\gtrsim$100~fs or $\gtrsim$100~ps \footnotemark[2] \\
Availability                               &high            &high             &low\footnotemark[3] \\
\end{tabular}
\footnotetext[1]{in connection with an assigned local mode and/or a specific IR label.}
\footnotetext[2]{free electron laser vs synchrotron, respectively.}
\footnotetext[3]{requires a large scale facility with limited access, such as a hard-X ray free electron laser or a synchrotron.}
\end{table*}

Light triggering is an attractive way to initiate non-equilibrium dynamics. \R{Such experiments require two ingredients: a structure-sensitive spectroscopic method with appropriate time resolution as well a photo-switchable component in the molecular system under study. As for the first, time-resolved IR spectroscopy (which is what we will focus on here, see Box 1) and time-resolved X-ray diffraction (XRD) are the methods of choice. Time-resolved XRD is based on synchrotrons, whose X-ray light is pulsed with a typical pulse duration of 100~ps,\cite{schotte03,Knapp2006} and more recently on hard X-ray free electron lasers.\cite{Kern2013,Nogly2018,Standfuss2019,Skopintsev2020} Table~\ref{tab1} compares the different methods that are used to study the structural dynamics of proteins. It might seem that time-resolved XRD is the one perfect method on this list, however, it requires the protein system to be crystalized, which very often is a bottleneck (X-ray free-electron lasers have relaxed that situation a bit since ``only'' nanocrystals are needed).\cite{chapman11}
Furthermore, the availability of such instruments is still very limited with 4-5 free electron lasers worldwide,  and very few time-points are typically measured. Time-resolved IR spectroscopy and time-resolved XRD thus complement each other. That is, the strength of time-resolved IR spectroscopy is the kinetics of a protein system (see e.g. Fig.~\ref{figTimeScale} below). Time-resolved XRD can then set in and determine the structures of various intermediates at selected time-points at which their population is maximal.

As for a photo-switchable component,} natural proteins use light for several purposes, most obviously for photosynthesis and vision, both of which have been studied extensively by various time-resolved techniques.\cite{wang94,zinth05,Palczewski2006,Engel2007,Kern2013,Mirkovic2017,Nogly2018} Furthermore, several photoreceptors exist that regulate the physiology of higher plants, microalgae, fungi and bacteria in response to environmental conditions.\cite{Hegemann2008,Rockwell2017} Myoglobin and hemoglobin are special in this regard: despite the fact that small molecule binding/unbinding, such as oxygen and CO, is not controlled by light in the natural system, one can still initiate these processes with light.\cite{Austin1975,schotte03,Knapp2006,cammarata2010}

\begin{figure}[t]
\begin{mdframed}
\vspace{0.5cm}
	\hspace{0.1\textwidth}\includegraphics[width=0.9\textwidth]{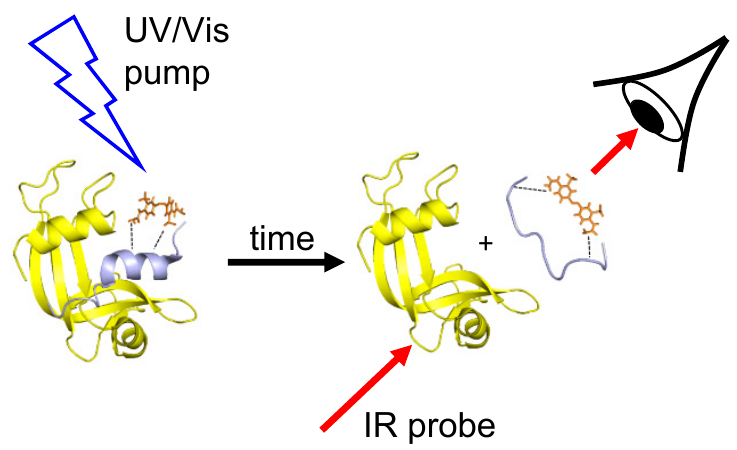}\\

\footnotesize \textbf{Box 1 Transient IR Spectroscopy}

In a transient IR experiment, one excites a photo-active protein with a short enough UV/Vis pump pulse at a proper wavelength, and probes its time-evolution with IR light. IR is the spectroscopy of choice when studying the non-equilibrium dynamics of proteins, since it features an intrinsic time resolution of 1~ps, fast enough to resolve any process that might be of biological relevance (in contrast to NMR spectroscopy), and since it also allows one to observe the response of the protein, and not only that of the photoactive cofactor (unlike transient UV/Vis spectroscopy). In particular the amide~I band around 1600-1700~cm$^{-1}$, in essence the C=O stretch vibrations of the protein backbone, is an extraordinarily sensitive reporter of protein structure.\cite{barth02} While it is impossible to determine a protein structure from the amide~I band, already very small conformational changes of a protein lead to measurable signals in the amide I spectrum and one can extract the timescales, on which these changes occur. Fig.~\ref{figTimeScale}a,c shows a typical result of such an transient IR experiment. Site-specific information can be obtained by incorporating IR labels with special molecular groups.\cite{koziol15}

\hspace{0.3cm} Technologies to measure transient IR spectra include step-scan FTIR spectroscopy,\cite{Murphy1975, uhmann91,Gerwert1999,Radu2009} quantum cascade lasers,\cite{Ritter2015,Zhang2014b,Schultz2018,Stritt2020,Klocke2018} as well as femtosecond laser based techniques.\cite{ham00b,Bredenbeck2004,Greetham2016,Jankovic2021b} Combining techniques, one can cover all relevant timescales from 100~fs to seconds and beyond.\\
\end{mdframed}
\end{figure}

The vast majority of biological processes, however, work without light. Indirect photocontrol can be achieved with caged compounds,\cite{Davies2007} pH jumps,\cite{causgrove06,donten15} or laser induced temperature jumps.\cite{thompson97,Callender98,Yang03,Hauser2008} On the other hand, Nature's strategy of using light for controlling biochemical processes may  serve as an inspiration for the more direct design of artificially photocontrollable proteins by involving photoswitchable molecules (or photoswitches for short). Photoswitches are small molecules that undergo a fast light-induced isomerization.\cite{Crespi2019, Lomas2012,Szymanski2013} By properly incorporating them into a protein system, one may modulate its function or activity directly,\cite{Willner1991,James2001, Liu1997, Hamachi1998,Yamada2007,Schierling2010,Ritterson2013,buchli13,Borowiak2015,Hoersch2016,Blacklock2018,Bozovic2020b,Bozovic2021} or one may manipulate the binding affinity of a photoswitchable ligand that binds to a protein.\cite{Guerrero2005, Guerrero2005b,Woolley2006,Kneissl2008,Zhang2010,Nevola2013,Martin-Quiros2015,Babalhavaeji2018,Albert2019a,Jankovic2019,Myrhammar2020,Bozovic2020a, Day2020,Jankovic2021,Jankovic2021b} Unlike naturally abundant photo-responsive proteins, artificial photocontrol allows for a rational design, depending on the effect one wants to achieve, which can be applied to virtually any protein system. Photoswitches are versatile and fully customizable, and the induced effects are often larger than those achieved with e.g. temperature or pH jumps.

\begin{figure*}[t]
	\begin{mdframed}
		\vspace{0.5cm}
		\includegraphics[width=1\textwidth]{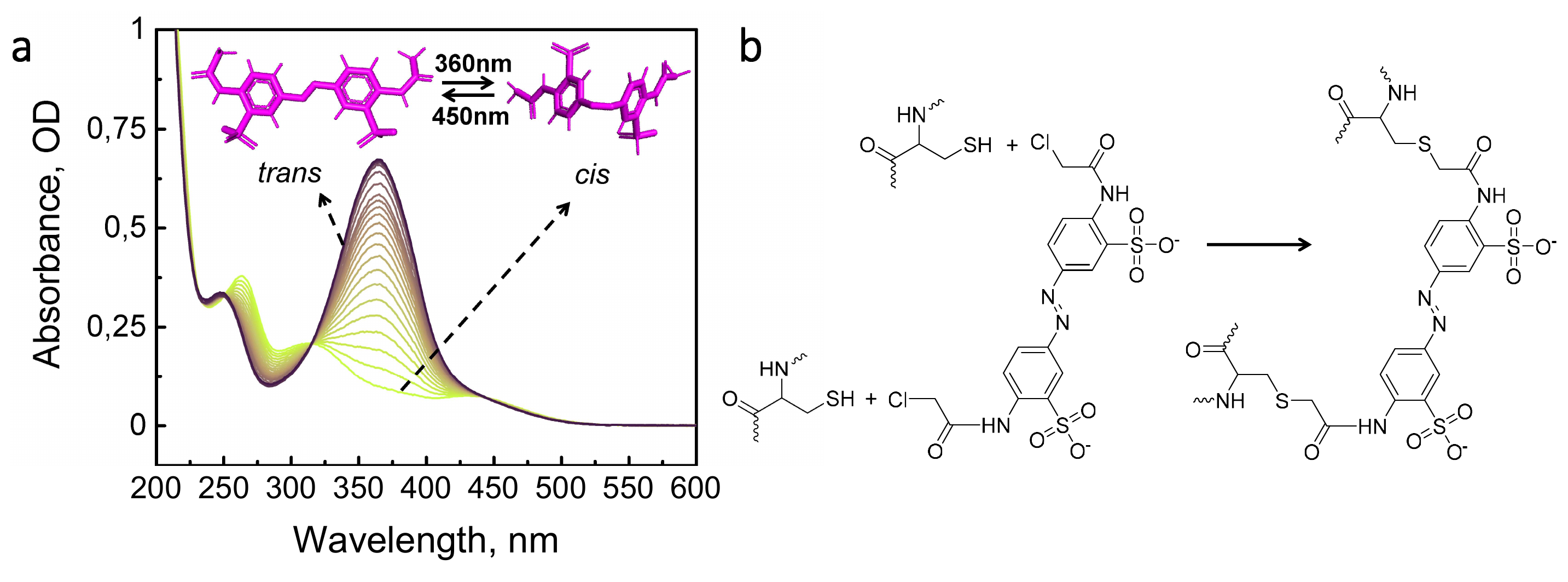}\\
		
\footnotesize \textbf{Box 2 Azobenzene-based photoswitches}
 		
Azobenzene-based molecules are the most widely used molecular photoswitches for controlling biological function.\cite{Beharry2011} Unlike many other approaches, azobenzene photoswitches undergo a reversible photoisomerisation upon illumination, and many cycles of photoswitching can be achieved. The UV/Vis absorption spectra of the \textit{trans} and \textit{cis} isomers are different from each other (panel a), most prominently at \mbox{ $\approx$ 360 nm} with the strong \mbox{$\pi$-$\pi$*} transition for the \textit{trans}-state (the exact position of the absorption maximum varies a bit depending on azobenzene derivative), and essentially no absorption in the \textit{cis}-state.
In contrast, the \mbox{n-$\pi$*} transition around 450~nm exists for both isomers.\cite{Beharry2011}

\hspace{0.3cm} The \textit{trans-}isomer of azobenzene is by \mbox{$\approx$ 10 kJ/mol} more stable than the \textit{cis-}isomer and therefore dominates if the molecule is kept in the dark.\cite{rau1973} Upon illumination of the \mbox{$\pi$-$\pi$*} transition, the photoswitch undergoes a \textit{trans}-to-\textit{cis} isomerisation, eventually resulting in a \textit{cis}-sample with purities of \mbox{$\geqslant$85\%}. The \textit{trans-}sample can then be re-established by keeping it in the dark and allowing for the equilibrium to be restored by thermal back-isomerisation, which happens on a minute to hour timescale. The sequence of spectra shown in the figure represents this process, each taken at a different time after preparing a \textit{cis}-sample.

\hspace{0.3cm} Time-resolved experiments with azobenzene controlled systems can be performed in both
directions. For \textit{trans}-to-\textit{cis} switching, one starts from a dark-adapted sample and excites either the \mbox{$\pi$-$\pi$*} or the \mbox{n-$\pi$*} transition of  \textit{trans}-azobenzene with a short laser pulse. For \textit{cis}-to-\textit{trans} switching, one first accumulates the \textit{cis}-state by pre-illuminating the \mbox{$\pi$-$\pi$*} transition of \textit{trans}-azobenzene, and then  excites the \mbox{n-$\pi$*} transition of  \textit{cis}-azobenzene with a short laser pulse. For time-resolved experiments, it is important to have only one photoswitch per protein. If more than one are used, one could not guarantee that all switch with the one pump laser pulse, since the excitation probability and the isomerisation quantum yield are smaller than 1.

\hspace{0.3cm} \R{Woolley and coworkers have pioneered a chemical concept,\cite{kum00} with which azoebenzene photoswitches can selectively be linked to two cysteines of a peptide or a protein in a simple, one-step reaction (panel b). Derivatives of the azobenzene photoswitch have been designed with additional -SO$_3^-$ groups that render it water-soluble.\cite{Zhang03}  }
	\end{mdframed}
\end{figure*}

Owing to many attractive features, the most widely used class of photoswitches are derivatives of azobenzene (see Box 2).\cite{Beharry2011} They are relatively small molecules, chemically very stable, and isomerize around the central \mbox{N=N} bond with high quantum yield.\cite{borisenko05}  The optical spectra of the two isomers, \textit{cis} and \textit{trans}, differ sufficiently to be able to steer the molecule to both states with high purity. The light-induced isomerisation proceeds through a conical intersection without any barrier, and is thus a very fast sub-picosecond process,\cite{naeg97} which imposes no limitation on the time-resolution of a transient experiment. Depending on substituents, the physical properties of azobenzene molecules may vary significantly. One can choose an appropriate molecule with excitation wavelength ranges between 300 and 700~nm,\cite{Dong2015,Dong2017} different solubility,\cite{Zhang03} and tunable relaxation rates.\cite{Beharry2011,sadovski2009} Many reviews focus on the application of azobenzene-based chemical modification of peptides and proteins, and here we refer to some of the more recent ones.\cite{Beharry2011,broichhagen2015,Ankenbruck2018,Zhu2018,hull2018,Albert2019,Paoletti2019} Other light reactive molecules have been explored as photoswitches as well, such as stilbene derivatives, hemithioindigo-based molecules, overcrowded alkenes and spiropyran-based compounds.\cite{Peddie2019} Nevertheless, azobenzene-based photoswitches remain the most widely used ones, \R{due to their superior photophysical properties, the simplicity of the photoreaction, their noninvasivness, reversibility, stability, and the ease of chemical modification.}\cite{hull2018}

\R{A major motivation to incorporate azobenzene photoswitches into proteins is to photocontrol biological activity. Examples reported in literature range from the regulation of enzymatic activity,\cite{Willner1991,James2001,Liu1997,Hamachi1998,Schierling2010, Hoersch2016, Blacklock2018, Yamada2007, hull2018} ion-channels and receptors,\cite{Paoletti2019, hull2018} all the way to cytoskeleton regulation,\cite{Borowiak2015}  cell-cell adhesion,\cite{Ritterson2013} as well as  \textit{in vivo} studies of various physiological processes.\cite{Beharry2011a,Zhang2010,Nevola2013,Borowiak2015}} But azobenzene photocontrol can also serve as a trigger to set proteins in motion, and this is where the full potential of this type of photocontrol comes in place. Exploiting the practically instantaneous isomerization of azobenzene photoswitches incorporated into proteins and peptides, and combining it with  time-resolved spectroscopic techniques, can reveal what happens to proteins once they are pushed out of the equilibrium. This combination will be the main topic of the present review article. In addition, we will discuss relevant applications of photocontrollable systems and how rational design can be used to modulate biological structure and function. Ultimately, we will reconcile the two aspects of protein dynamics and show how fundamental questions can be addressed, which require a time dimension.

\begin{figure}[t]
	\centering
	\includegraphics[width=0.45\textwidth]{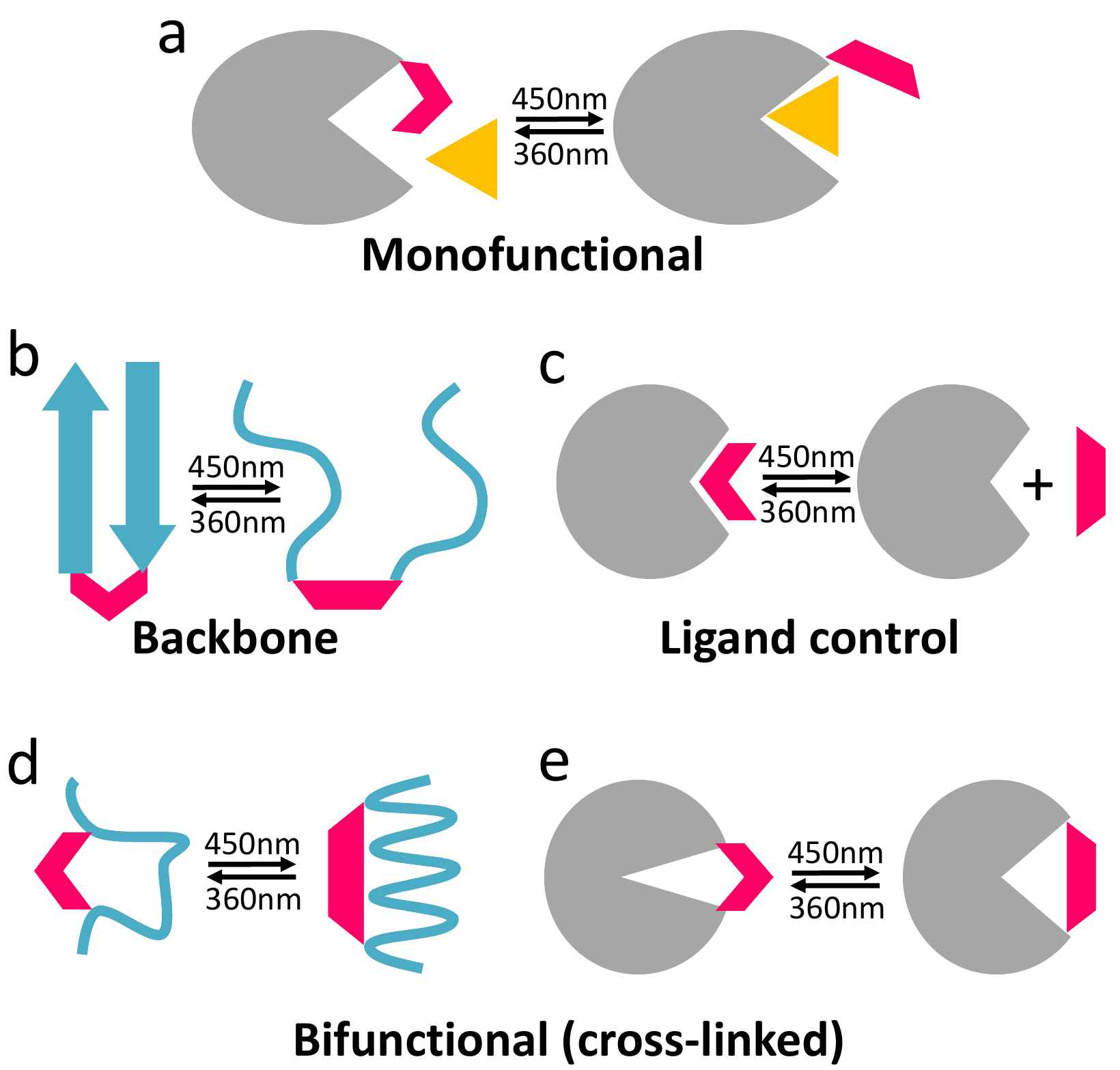}
	\caption{Various strategies of azobenzene photocontrol: (a) monofunctional control, (b) photoswitch incorporated into the backbone, (c) ligand switching, as well as bifunctional (d) $\alpha$-helix switching and (e) cross-linking different secondary structure motifs of a protein. } \label{figPhotoswitching}
\end{figure}

\section{Strategies of Azobenzene Photocontrol}

\R{The first realisations of photoswitchable proteins linked the azobenzene moiety to a single site,}\cite{Willner1991,Paoletti2019,Yamada2007} or incorporated an azobenzene-moiety bearing amino acid phenylazophenylalanine.\cite{Liu1997,Hamachi1998,James2001} This is called ``monofunctional control'' and illustrated in Fig.~\ref{figPhotoswitching}a. The azobenzene-moiety interferes sterically with the natural agonist and thereby modulates the activity of the protein system. Furthermore, azobenzene derivatives mimicking an amino acid, with a carboxyl group on one ring and an amino group on the other, have been incorporated directly into a peptide backbone by modified peptide synthesis, in essence introducing a kink in the backbone upon \textit{trans}-to-\textit{cis} isomerisation (Fig.~\ref{figPhotoswitching}b), and thereby controlling the fold of the peptide.\cite{behr99,spo02,bre03,aemissegger05,rehm05,schrader07,Rampp2018,Day2020}

The applicability of azo-switches increased tremendously with works by Woolley and coworkers, who introduced a very versatile concept, post-translationally cross-linking two sites of a protein by a photoswitch (bifunctional control, see Figs.~\ref{figPhotoswitching}d,e).\cite{kum00} To that end, the selective reaction of the thiol side-chains of two cysteines of a protein with a properly designed azobenzene derivative is utilized \R{(see Box2)}. By site-directed mutagenesis, these two cysteines can be placed at virtually any position of a protein, opening a huge playground of applications.

Most applications of the approach target $\alpha$-helices (Fig.~\ref{figPhotoswitching}d), either isolated\cite{flint02,woolley05,chen03,bre05a,ihalainen07,Ihalainen2008} or as secondary structure motifs in larger proteins.\cite{Nevola2013, Jankovic2019, Jankovic2021,Jankovic2021b,Jankovic2019,Kneissl2008, Myrhammar2020, Zhang2010, Guerrero2005, Guerrero2005b, Woolley2006,Bozovic2020b,Bozovic2021,Martin-Quiros2015,Schierling2010} The light-induced isomerization of an azobenzene molecule around the \mbox{N=N} bond changes the distance between the two anchoring points, and thus controls the $\alpha$-helical content mechanically. Woolley and coworkers have shown that the helical content is stabilized in the \textit{cis}-state of the azobenzene relative to that in the \textit{trans}-state, when the spacing between the two anchoring points of the photoswitch is $<$9, while it is destabilized for larger spacings (Fig.~\ref{figPhotoswitching}d).\cite{flint02,woolley05}

The concept of azobenzene photoswitching is in fact much more versatile, as one can control any secondary structure element, such as $\beta$-sheets,\cite{Bozovic2020a} or loops.\cite{Schierling2010, Ritterson2013} Furthermore, two secondary structure motifs within a protein have been linked by a photoswitch in order to modulate their relative distance and/or orientation (Fig.~\ref{figPhotoswitching}e).\cite{buchli13,Schierling2010,Hoersch2016,Zhang2009,Babalhavaeji2018,Ritterson2013, Blacklock2018} Control of biological activity has also been achieved via photocontrollable small molecule ligands or inhibitors that interact with a target protein (Fig.~\ref{figPhotoswitching}c).\cite{hull2018,Zhu2018,Borowiak2015}

\section{Time-resolved Studies on Small Peptides: Simple Model Systems}

Time-resolved studies of azobenzene controlled biomolecules are quite scarce, and began with investigating the folding dynamics of small peptides. In a time-resolved experiment, one switches the azobenzene photoswitch either from the \textit{cis}- to the \textit{trans}-state, or \textit{vice versa}, and observes the response of the photoswitch and the peptide by transient Vis or transient IR spectroscopy, see Box 1.
In the first attempt to understand the conformational dynamics of azobenzene-based peptides, Ref.~\onlinecite{spo02} employed a small cyclic peptide with the photoswitch embedded in the peptide backbone. The ultrafast time-resolved response was followed in the UV/Vis spectral region, using the azobenzene molecule not only as the initiator of a perturbation, but also as a spectroscopic observable. The study showed that the majority of conformational rearrangement of the photoswitch happens within the first 50~ps after excitation. In follow-up work on the the same cyclic peptide, the peptide backbone response has been monitored employing transient IR spectroscopy of the amide~I band.\cite{bre03} This enabled the observation of the subsequent peptide relaxation, which extends up to 16~ns.

With the help of time-resolved optical rotatory dispersion, Ref.~\onlinecite{chen03} studied the folding dynamics of a 16 amino acid long $\alpha$-helical peptide. By cross-linking the photoswitch between two cysteines separated by 11 amino acids, the system had a higher helical propensity in the \textit{trans}-state of the photoswitch, while isomerisation to \textit{cis}-state lead to disruption (unfolding) of the helix (see Fig.~\ref{figPhotoswitching}d). It was seen that the forced unfolding of the helix occurs within 55~ns. The same $\alpha$-helical construct was also investigated by transient IR spectroscopy. Taking advantage of the photoswitching reversibility, both the folding and unfolding direction has been investigated,\cite{ihalainen07} and that in a site-selective manner with the help of C$^{13}$=O$^{18}$ isotope labelling.\cite{Ihalainen2008} The typical timescale of helix folding is significantly slower, 300~ns to 3~$\mu$s depending on temperature.

The folding and unfolding of a tryptophan zipper analogue containing a $\beta$-hairpin has been addressed as well with the construct shown in Fig.~\ref{figPhotoswitching}b.\cite{schrader07,Rampp2018} Photoswitching initiated a conformational transformation of the system between a $\beta$-hairpin and an unfolded hydrophobic cluster. Unfolding takes a few nanoseconds only, while the reverse folding of this system requires more sampling of the conformational space and occurs on a 30~$\mu$s timescale.

The works on small peptides not only contributed to a better understanding of the protein folding problem, but also paved the way for subsequent studies on more complex artificially photocontrollable protein systems, which will be discussed in the next Chapter.

\begin{figure}[t]
	\centering
	\includegraphics[width=0.4\textwidth]{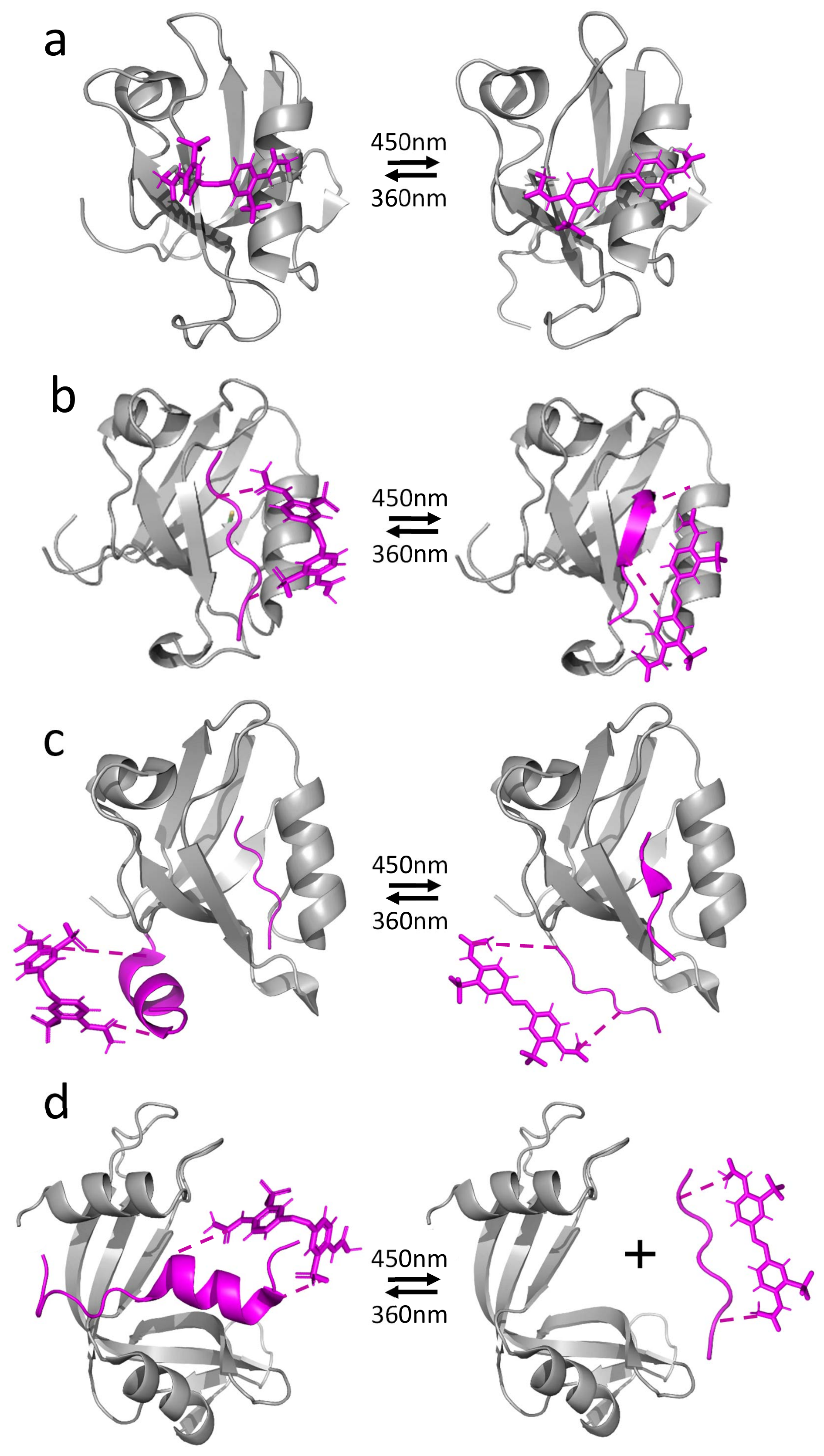}
	\caption{Various artificially photoswitchable proteins that have been studied in the context of protein allostery and protein-ligand binding by time-resolved spectroscopy. (a) PDZ2 domain with the binding-groove cross-linked by a photoswitch,\cite{buchli13,waldauer14} (b) PDZ2 domain with a photoswitchable peptide ligand,\cite{Bozovic2020a} (c) PDZ3 domain with a photoswitchable $\alpha3$-helix\cite{Bozovic2020b,Bozovic2021} and (d) the RNase~S complex with a photoswitchable S-peptide.\cite{Jankovic2019,Jankovic2021,Jankovic2021b}} \label{figMolecules}
\end{figure}

\begin{figure*}[t]
	\centering
	\includegraphics[width=0.8\textwidth]{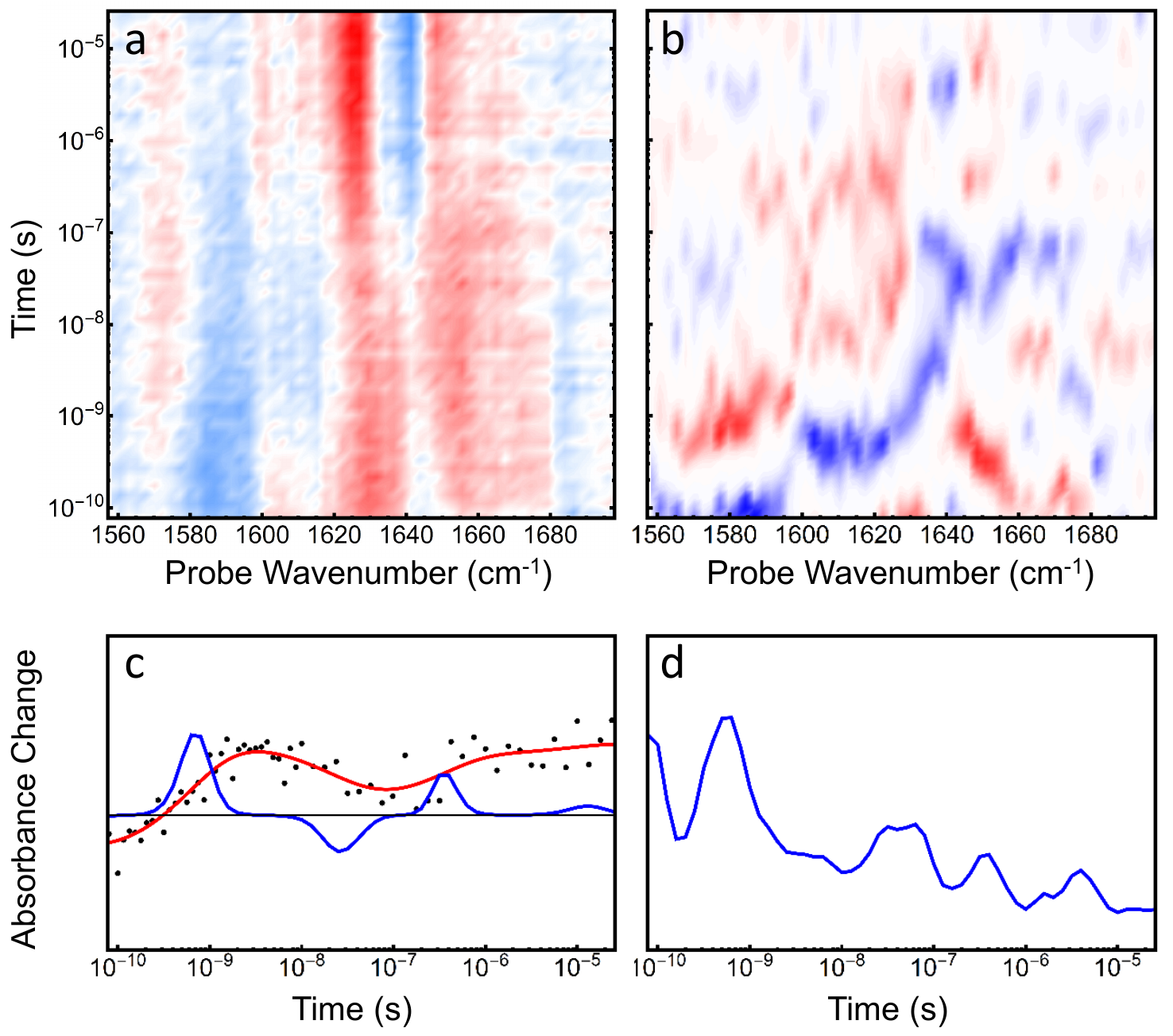}
	\caption{(a) Transient IR data upon \textit{cis}-to-\textit{trans} switching of the PDZ2 system shown in Fig.~\ref{figMolecules}b and (b) its lifetime spectrum obtained from fitting it to Eq.~\ref{EqMultiexp}. Positive amplitudes are plotted in red, negative amplitudes in blue. (c) Transient signal exemplified for probe position  1643~cm$^{-1}$ with its fit (red) and the corresponding lifetime spectrum (blue). (d) Averaged dynamical content calculated according to Eq.~\ref{Eqdyncontent}. Adapted with permission from Ref.~\onlinecite{Bozovic2020a}.} \label{figTimeScale}
\end{figure*}

\section{Time-Resolved Studies of Protein Systems}

Larger proteins typically do not completely fold/unfold upon photoswitching, since the configurational change of the azobenzene moiety is too small to initiate such a large transition (Ref.~\onlinecite{Zhang2009a} is an exception in this regard). Rather, one changes a local structure element and thereby the function of the protein system. There are two approaches to artificially photocontrol a protein and subsequently investigate it with time-resolved methods. First, one can decide to photocontrol a binding partner for the system in question (Fig.~\ref{figMolecules}b,d).\cite{Jankovic2019,Bozovic2020a,Jankovic2021,Jankovic2021b} These partners usually are relatively short peptide ligands, which can be synthesized using standard solid state peptide synthesis and subsequently cross-linked with an azobenzene molecule. This approach is significantly less demanding than the second one,  where a photoswitch molecule is inserted postranslationally to a full length protein(Fig.~\ref{figMolecules}a,c).\cite{buchli13,waldauer14,Bozovic2020b,Bozovic2021} Deciding on the method depends on the general properties of the system as well as the scientific questions posed. Furthermore, one needs to keep in mind that proteins need water as solvent, in contrast to the peptide systems discussed in the previous Chapter. This typically requires a photoswitch that is made water-soluble by adding polar groups to the azobenzene rings.\cite{Zhang03}

Time-resolved studies of artificially photocontrollable proteins focused on two classes of proteins, PDZ domains (Fig.~\ref{figMolecules}a-c) and the RNase~S complex (Fig.~\ref{figMolecules}d). PDZ domains are ubiquitous protein interaction domains found in a wide range of organisms from bacteria to mammals (Postsynaptic density protein/Drosophila disc large tumour suppressor/Zonula occludens-1 protein).\cite{Lee2010b,Ivarsson2012} PDZ domains are relatively small protein domains that consist of approximately 100 amino acids.\cite{MoraisCabral1996} They share a common fold and are usually integral parts of larger, multidomain proteins.\cite{basdevant2006,munz2012,gerek2009,Tonikian2008,TeVelthuis2011} Given that PDZ domains are an important class of protein interaction domains, they have been extensively studied in the context of protein allostery.\cite{petit09,swain2006,Fuentes06}  Allosteric communication in PDZ-containing proteins was shown to be possible between coupled domains, so-called supertertiary structures, as well as within single, isolated domains.\cite{Li2009,Fuentes2004,Fuentes06} The size of PDZ domains, the ease of production and manipulation, as well as their biological omnipresence and importance, make isolated PDZ domains a worthwhile target for time-resolved studies.

The \mbox{RNase~S} complex (Fig.~\ref{figMolecules}d), on the other hand, is a product of the selective cleavage of RNase A by the enzyme subtilisin, which cleaves a specific single peptide bond.\cite{richards1958} RNase A is an enzyme that catalyzes the hydrolysis of a phosphodiester bond in \mbox{RNA}.\cite{roberts1969} The \mbox{RNase S} complex consists of the S-protein and the shorter S-peptide fragment. The isolated S-protein does not possess any catalytic activity, given that the second histidine residue necessary for hydrolysis of RNA is located on the S-peptide. Nevertheless, when co-dissolved, S-protein and S-peptide re-associate with nanomolar affinity and restore the full enzymatic activity.\cite{richards1959} Another peculiarity of this system lies in the fact that the isolated S-protein preserves its folded structure to a certain extent, as evidenced from circular dichroism spectra (no NMR spectra or crystal structures are available), while the isolated S-peptide is an unstructured random coil. Upon re-association of the S-peptide with the S-protein, the RNase~S complex restores a structure that is practically the same as that of \mbox{RNase A},\cite{wlodawer82} with the S-peptide being $\alpha$-helical. The RNase~S system is a widely used model system to study protein-peptide binding mechanisms, in particular in the context of ``induced fit'' vs ``conformational selection''.\cite{Goldberg1999, Bachmann2011, Luitz2017, Schreier1977}

Utilizing these two protein systems as examples, we will discuss in the following how proteins respond to a perturbation induced by a photoswitch, determine the speed of an allosteric signal, and study the full sequence of events during ligand unbinding.

\begin{figure}[t]
\begin{mdframed}
\vspace{0.5cm}
\footnotesize\textbf{Box 3 Timescale Analysis and Dynamical Content}

The timescales contained in a transient data set such as that of Fig.~\ref{figTimeScale}a are determined by fitting each kinetic trace at probe-frequency $\omega_i$ to a multiexponetial function:
\begin{equation}
  S(t,\omega_i) = a_0- \sum_k a(\omega_i,\tau_{k})e^{-t/\tau_{k}} \label{EqMultiexp},
\end{equation}
where a maximum entropy method has been applied for a regularisation of the otherwise ill-poised inverse Laplace transformation.\cite{Hobson1998,Kumar2001,Lorenz-Fonfria2006} In this fit, the timescales $\tau_{k}$ were fixed and equally distributed on a logarithmic scale with 10 terms per decade, while the amplitudes $a(\omega_i,\tau_k)$ are the free fitting parameters. In Ref.~\onlinecite{Buhrke2020}, we have compared this approach to  more conventional global fitting. The amplitudes $a(\omega_i,\tau_k)$ are called ``timescale spectra'' or sometimes ``dynamical content''.\cite{Shaw2010}

\hspace{0.3cm} Fig.~\ref{figTimeScale}c exemplifies the concept for one particular probe frequency $\omega_i$. On a logarithmic time axis, an exponential process exhibits a sigmoidal shape. Whenever such a sigmoidal step occurs in the data, which can be up or down, the corresponding lifetime spectrum shown in blue reveals a peak. Fig.~\ref{figTimeScale}b assembles the lifetime spectra $a(\omega_i,\tau_k)$ for all probe frequencies into a 2D plot, illustrating the reach information content of the data.

\hspace{0.3cm} Provided that the underlying dynamics can be understood in terms of a Markov State Model (MSM, see Box 4), all kinetic traces contain, in principle, the same implied timescales, albeit with different amplitudes. To extract these timescales, the timescale spectra $a(\omega_i,\tau_{k})$ are averaged over all probe frequencies $\omega_i$, taking into account that they can have positive or negative signs:\cite{Stock2018}
\begin{equation}
  D(\tau_{k})=\sqrt{\sum_i a(\omega_i,\tau_{k})^2}. \label{Eqdyncontent}
\end{equation}
The averaged dynamical content encompass all kinetic processes of a data set, and avoids that one has to hand-pick kinetic traces at particular probe frequencies. Fig.~\ref{figTimeScale}d shows that a set of discrete time scales still exists in the averaged dynamical content, which we attribute to the implied timescales of the underlying MSM, i.e., the relaxation on the rugged energy landscape of the protein.\\
\end{mdframed}
\end{figure}

\begin{figure*}[t]
\begin{mdframed}
\vspace{0.5cm}
	\hspace{0.15\textwidth}\includegraphics[width=0.8\textwidth]{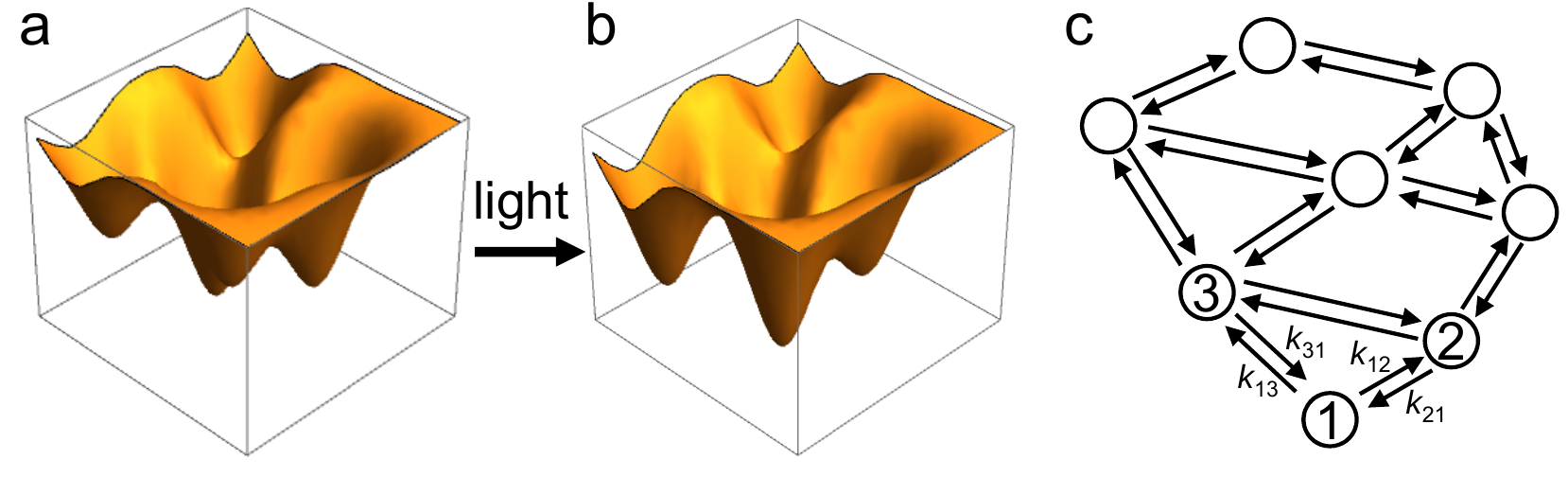}\\
\footnotesize\textbf{Box 4 Rugged Energy Landscape and Markov State Models}

Panels (a) and (b) sketch the rugged energy landscape of a protein.\cite{Frauenfelder91,Dill1997} That is, proteins do not adopt a single rigid structure, rather an ensemble of closely related, but not identical structures, each of which characterized by a local minimum of the free energy surface of the protein. Consider for example several structures of a floppy loop region of a protein. According to an emerging new view, the allosteric signal is related to a remodelling of the free energy landscape, i.e., the depths of the various free energy minima change upon an allosteric signal (compare panels a and b), and thereby shift the populations between substates.\cite{swain2006,Tsai2008,smock2009,Tsai2014,Hilser2012,Bozovic2020a}

\hspace{0.3cm} The kinetic consequence of that picture can be understood in terms of a Markov State Model (MSM),\cite{Bowman2013,pande10,prinz11,Sengupta2018,Bozovic2020a} which in essence describes a set of states connected by microscopic rates $k_{ij}$ (see panel c) that are determined by the barriers connecting the local minima $i$ and $j$. The corresponding kinetic equation is:
\begin{equation}
\frac{d \boldsymbol{p}}{dt}=\boldsymbol{K}\cdot \boldsymbol{p}, \label{eqMarkov}
\end{equation}
where $\boldsymbol{p}$ is a vector whose components $p_i$ contain the populations of state $i$, and $\boldsymbol{K}$ a kinetic matrix:
\begin{equation}
\boldsymbol{K}=
\begin{pmatrix}
-k_{12}-k_{13} & k_{21} & k_{31} & \dots\\
k_{12}         & \ddots &      & \\
k_{13}         &      & \ddots & \\
\vdots         &&&\ddots
\end{pmatrix},
\end{equation}
exemplifying here only the microscopic rates shown in panel (c). The free-energy difference $\Delta F_{ij}$ between pairwise two states $i$ and $j$ is related to the microscopic rates by detailed balance:
\begin{equation}
\frac{k_{ij}}{k_{ji}}=e^{-\frac{\Delta F_{ij}}{k_BT}}. \label{eqfreeenergy}
\end{equation}
The kernel (nullspace) of the kinetic matrix $\boldsymbol{K}$, setting the left-side of Eq.~\ref{eqMarkov} to zero, determines the population $\boldsymbol{p}_{eq}$ in equilibrium. Initially, the protein will be in equilibrium with respect to a kinetic matrix $\boldsymbol{K}_b$ (where the subscript \textit{b} stands for ``before''). Upon an allosteric signal (here upon light-triggering), the free energy surface changes and correspondingly also the kinetic matrix $\boldsymbol{K}_a$ (for ``after''), along the lines of Eq.~\ref{eqfreeenergy}. Immediately after the trigger, the populations will still be sameas before, hence the system all of the sudden will be out of equilibrium. It will subsequently equilibrate according to Eq.~\ref{eqMarkov} with the kinetic matrix $\boldsymbol{K}_a$.

\hspace{0.3cm} The dynamical content (see Box 3) of a transient experiment will \textit{not} contain the microscopic rates $k_{ij}$, but rather the eigenvalues of the  kinetic matrix $\boldsymbol{K}_a$,\cite{prinz11} which are sometimes called the ``implied timescales'' of the MSM. The number of implied timescales equals the number $n$ of states in the MSM minus 1, which is a significantly smaller number than that of  microscopic rates (of the order $n^2$). The relatively small number of implied timescales facilitates the discreteness of the dynamical content shown in Fig.~\ref{figTimeScale}d. It is important to stress that the same set of implied timescales is revealed, regardless of the observable one is looking at, but different observables may weigh the various implied timescales differently. For example, each probe frequency in Fig.~\ref{figTimeScale}b reveals a different response, yet the averaged dynamical content in Fig.~\ref{figTimeScale}d emphasizes that there is a common set of timescales.\\
\end{mdframed}
\end{figure*}

\subsection{Protein Response}

As a first step towards understanding the non-equilibrium allosteric nature of proteins, Ref.~\onlinecite{buchli13} presented a PDZ2 domain, in which the photoswitch was linked across its binding groove (Figs.~\ref{figPhotoswitching}e and \ref{figMolecules}a). Isomerization of the photoswitch forces the opening or closing of the binding groove. Albeit an artificial construct by default, it was shown by solving NMR structures for the photoswitchable protein in the \textit{cis}- and the \textit{trans}-state that the perturbation closely mimics the structural change of a native PDZ2 domain upon ligand binding/unbinding. Transient infrared spectroscopy revealed three major phases of the overall process. Initial photo-excitation is followed by fast photoisomerization and heat dissipation on a 10~ps timescale, an effect widely seen in these type of experiments.\cite{ham97b, Baumann2019} Subsequently, the photoswitch relaxes structurally on a 10~ns timescale, as judged from a vibrational mode that is localized on the linker between the azobenzene moiety of the photoswitch and the protein backbone. The structural relaxation slows down when attaching the photoswitch to the protein, owing to the strain imposed by the protein, which cannot accommodate the structural change of the photoswitch immediately. This phase was thus assigned to the perturbation of the binding groove of the PDZ2 domain. Finally, the perturbation of the binding groove propagates through the whole protein within 10~$\mu$s, as deduced from the amide~I band, which is an extraordinarily sensitive reporter of protein structure.\cite{barth02}

Exactly this propagation of a perturbation through the protein was investigated more closely in a follow-up study,\cite{Bozovic2020a} utilizing a less artificial PDZ2 construct (Fig.~\ref{figMolecules}b). That is, the PDZ2 domain was kept intact in this case, while a newly introduced peptide ligand was made photoswitchable. \textit{Cis}-to-\textit{trans} isomerisation of the peptide ligand stabilizes its $\beta$-strand structure in the binding groove of the PDZ2 domain and thus increases its binding affinity by 5 fold.  It also injects an allosteric signal into the PDZ2 domain, that has been investigated by transient infrared spectroscopy in connection with isotope labelling of the whole protein. Fig.~\ref{figTimeScale}a shows a typical example of such an experiment, plotting the absorption change of the protein at various probe frequencies within the amide~I band as a function of time; the latter on a logarithmic scale to be able to cover a wide range of timescales. Fig.~\ref{figTimeScale}b shows the corresponding timescale analysis (see Box 3 for an explanation), emphasizing the very rich information content of such an experiment.

Molecular dynamics (MD) simulations were performed to understand this complex response.\cite{Bozovic2020a} They revealed that PDZ2 exists in a couple of conformational sub-states, in essence reflecting what is known as the ``rugged energy landscape'' of proteins.\cite{Frauenfelder91,Dill1997} Photoswitching of the ligand redistributes the population between those sub-states to a relatively small extent, in accordance with an emerging new view of allostery.\cite{swain2006,Tsai2008,smock2009,Tsai2014,Hilser2012} The overall change of protein structure is only 0.3~\AA, illustrating the extraordinary sensitivity of transient IR spectroscopy, which is able to detect such small changes.

The kinetic consequence of the rugged energy picture can be understood in terms of a Markov State Model (MSM), which is explained in more detail in Box 4. Most importantly, one expects a relatively small number of discrete timescales, the ``implied'' timescales of the MSM, as indeed observed in the averaged dynamical content of the transient IR data shown in Fig.~\ref{figTimeScale}d. MD simulations of the first PDZ2 construct discussed above (Fig.~\ref{figMolecules}a) revealed qualitatively the same behaviour,\cite{Buchenberg2017,Stock2018} and we have also observed it in a different PDZ domain\cite{Bozovic2021} as well as in a naturally photoswitchable protein.\cite{Buhrke2020} It thus appears to be an universal property of photoswitchable proteins.

\begin{figure*}[t]
	\centering
	\includegraphics[width=0.8\textwidth]{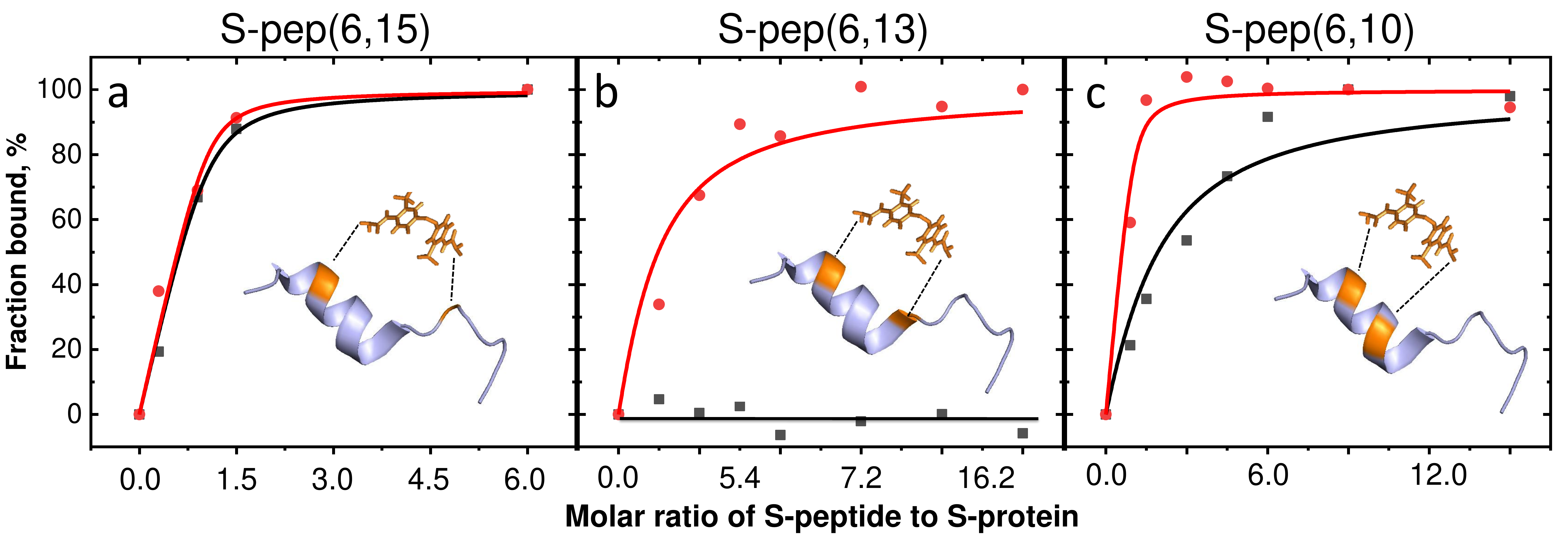}
	\caption{Binding curves for the \textit{cis} (red) or \textit{trans}-states (black) of the S-peptide to the S-protein for (a) spacing 9 between the two anchoring points (S-pep(6,15), indicating the positions of the two anchoring points), (b) spacing 7 (S-pep(6,13)), and (c) spacing 4 (S-pep(6,10)). Adapted with permission from Ref.~\onlinecite{Jankovic2019}, Copyright 2019 American Chemical Society. } \label{figBinding}
\end{figure*}

\subsection{Allosteric Signal}

The PDZ3 domain differs from the rest of the PDZ family by an additional $\alpha$-helix at its  C-terminus.\cite{doyle1996,MoraisCabral1996} This helical extension is distant from the binding groove and was shown to be an allosteric element of the protein.\cite{ballif2008,zhang2011a,petit09}
In Ref.~\onlinecite{Bozovic2020b}, a photocontrollable PDZ3 protein was designed by cross-linking the photoswitch to that $\alpha$3-helix (Fig.~\ref{figMolecules}c). Isomerization of the photoswitch leads to the perturbation of the helical structure and changes the binding affinity for a peptide ligand up to 120-fold, depending on temperature. Moreover, the perturbation of the helix not only changes the affinity in the binding groove, but the binding of a ligand also speeds up the thermal \textit{cis}-to-\textit{trans} isomerization rate of the photoswitch, introducing the concept of an ``allosteric force''. This is allostery in its literal sense, as one may consider the \textit{cis}-to-\textit{trans} isomerization of the azobenzene moiety a chemical reaction. The PDZ3 domain thus is arguably the smallest fully allosteric protein with a well-defined allosteric and effector site.

The unique properties of the photoswitchable PDZ3 domain make it an ideal candidate to shed light on the very nature of the allosteric signal. Employing a combination of transient UV/Vis and IR spectroscopies, the propagation of the allosteric signal could be reconstructed.\cite{Bozovic2021} Similar to the kinetics of the forced unfolding of isolated helices discussed above,\cite{chen03,ihalainen07} the $\alpha$3-helix unfolds (partially) on a 5~ns timescale after photoswitching. The protein responds to that perturbation on multiple time-scales up to 10~$\mu$s, again representing the ruggedness of its free energy landscape. However, by double-difference spectroscopy, the response of the peptide ligand could be singled out, revealing one dominating timescale of 200~ns, which has been attributed to the speed of the allosteric signal within the protein.\cite{Bozovic2021}

\subsection{Ligand Unbinding}

Biomolecular recognition is a major mechanism of protein regulation. Protein ligand interactions are fundamental in any biochemical process and have been studied in great depth for a plethora of systems. The RNase~S complex (Fig.~\ref{figMolecules}d) represents one such system, which has been widely used as a model for studying coupled binding and folding of the S-peptide during the interaction with the S-protein. Various photoswitchable variants of a S-peptide have been designed with different spacings and positions of the anchoring points of the photoswitch, and their binding affinities to the S-protein have been measured with CD spectroscopy.\cite{Jankovic2019} If the spacing of the two anchoring points is 9 (S-pep(6,15)), photoswitching has virtually no effect on the binding affinity (Fig.~\ref{figBinding}a), in accordance with the observation for isolated helices that this spacing is the dividing point between the two regimes, in which either the \textit{cis} or the \textit{trans}-state stabilizes the helix.\cite{flint02,woolley05} For shorter spacings, the helical structure of the S-peptide is stabilized in the \textit{cis}-state of the photoswitch, facilitating binding to the S-protein. The binding affinity changes by more than 20 fold when the spacing is 4 (S-pep(6,10), see Fig.~\ref{figBinding}c), in the same order as observed for other protein constructs.\cite{Guerrero2005,Woolley2006,Kneissl2008} S-pep(6,13) with spacing 7 is unique in this regard, as it binds in the \textit{cis}-state with reasonable binding affinity, while no specific binding could be detected in the \textit{trans}-state (Fig.~\ref{figBinding}b). Molecular dynamics (MD) simulations confirmed that the underlying mechanism for the modulation of the binding affinity does lie in the disruption of the helical content of the S-peptide,\cite{Jankovic2019} setting off the possibility of a rational design of this type of molecular systems.

\begin{figure*}[t]
	\centering
	\includegraphics[width=0.95\textwidth]{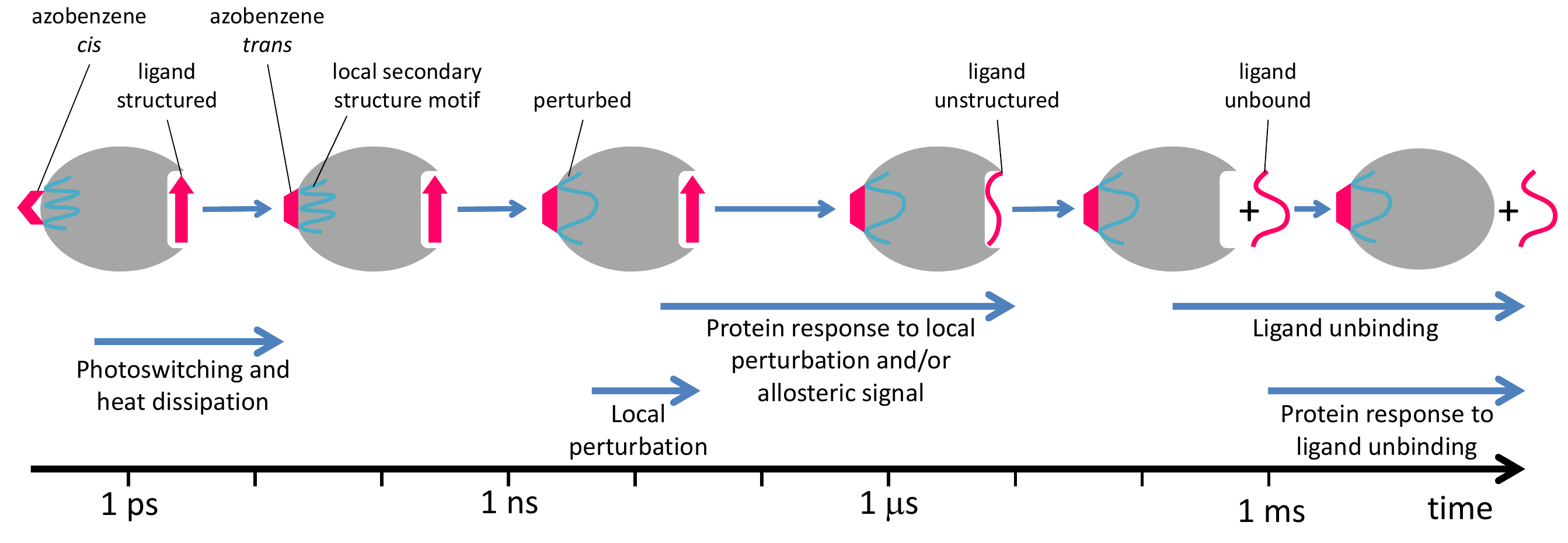}
	\caption{Full sequence of events of allosteric propagation and ultimate ligand unbinding together with their typical timescales.} \label{figConclusion}
\end{figure*}

The research of protein-ligand binding circles around the long-lasting debate of two limiting scenarios - induced fit vs conformational selection.\cite{boehr2009, morando2016, csermely2010} The necessary prerequisite for such a classification rests on the temporal sequence of binding event vs conformational changes of the binding partners.\cite{Vogt2012, Gianni2014, Hammes2009}
The on/off behaviour of S-pep(6,13) made it an interesting model system to study the complete sequence of events during the unbinding of the ligand by transient fluorescence and
IR spectroscopy.\cite{Jankovic2021, Jankovic2021b} Upon \textit{cis}-to-\textit{trans} switching, unbinding of the ligand is essentially a barrier-less process and proceeds in the fastest possible manner, revealing the ``speed limit'' of ligand unbinding. As such, it is less of a rate-limiting step and enables the observation of slower relaxation processes of the protein.

Ref.~\onlinecite{Jankovic2021b} uncovered the full sequence of events during unbinding. Transient IR spectroscopy revealed that the helical S-peptide, still inside the binding groove, unfolds within 20~ns; a similar timescale as that observed for isolated $\alpha$-helices\cite{chen03,ihalainen07} or for the $\alpha3$-helix in the PDZ3 system.\cite{Bozovic2021} The peptide remains bound, either in the binding pocket or to other parts of the protein, for quite some time and leaves only after about 300~$\mu$s; four orders of magnitudes later in time.\cite{Jankovic2021} The protein subsequently responds to that on a 3~ms timescale and an even slower  process that could not be observed with the experimental setup of Ref.~\onlinecite{Jankovic2021b}. Encompassing all the underlying events before, during and after the peptide unbinding, we offered a comprehensive explanation of the RNase S binding mechanism. That is, from the perspective of the S-peptide, induced fit seems to be the predominant mechanism, since it can explore its conformational space on timescales much faster (20~ns) than it remains bound to the protein, even for a ligand that has been designed to unbind as quickly as possible. To understand this argument, one has to keep in mind that the experiment of Ref.~\cite{Jankovic2021b} investigated peptide unbinding, while the induced fit scenario argues from a binding. On the other hand, the behaviour of the S-protein is better explained as conformational selection, as its conformational dynamics is slower than unbinding. These studies showcased the importance of the non-equilibrium approach to unequivocally discover the sequence of events on all relevant timescales and ultimately clarify the binding mechanisms.

\subsection{Typical Timescales}
Compiling the results obtained from the experiments we just discussed,\cite{buchli13,Bozovic2020a,Bozovic2021,Jankovic2021,Jankovic2021b} the picture shown in Fig.~\ref{figConclusion} emerges. The results have been obtained for different protein systems (Fig.~\ref{figMolecules}), not all of them revealing all processes shown in Fig.~\ref{figConclusion}. Nevertheless, the different protein systems share quite a few common features with respect to the type of response and their typical timescales, and we can put them into the unified representation of Fig.~\ref{figConclusion}.

The isomerisation process and the subsequent dissipation of the released heat universally occurs on a sub-nanosecond timescale. The photoswitch then perturbs the local structure element to which it is directly bound on a roughly 10~ns timescale, e.g. the opening of the binding groove of the PDZ2 construct shown in Fig.~\ref{figMolecules}a, or the partial unfolding of an $\alpha$-helix in the PDZ3 domain (Fig.~\ref{figMolecules}c) or the RNase~S complex (Fig.~\ref{figMolecules}d). The perturbation subsequently propagates through the whole protein on multiple timescales, ranging from $\approx$10~ns to $\approx$10~$\mu$s, a characteristics we have observed for all protein systems we have studied so far. If it is an allosteric protein, that process includes the allosteric signal (found to reach the binding groove in the case of PDZ3 system within 200~ns). Ligand unbinding occurs  only after a few 100~$\mu$s in the fastest possible cases, such as S-pep(6,13) bound to the S-protein, and the protein adapts to that change yet another time on a few millisecond timescale and beyond. To the best of our knowledge, this is the most comprehensive picture as of today of what is happening inside a protein during allosteric signalling. It explains how proteins adapt and respond to an allosteric signal, which ultimately leads to ligand unbinding.

\begin{figure*}[t]
	\centering
	\includegraphics[width=0.9\textwidth]{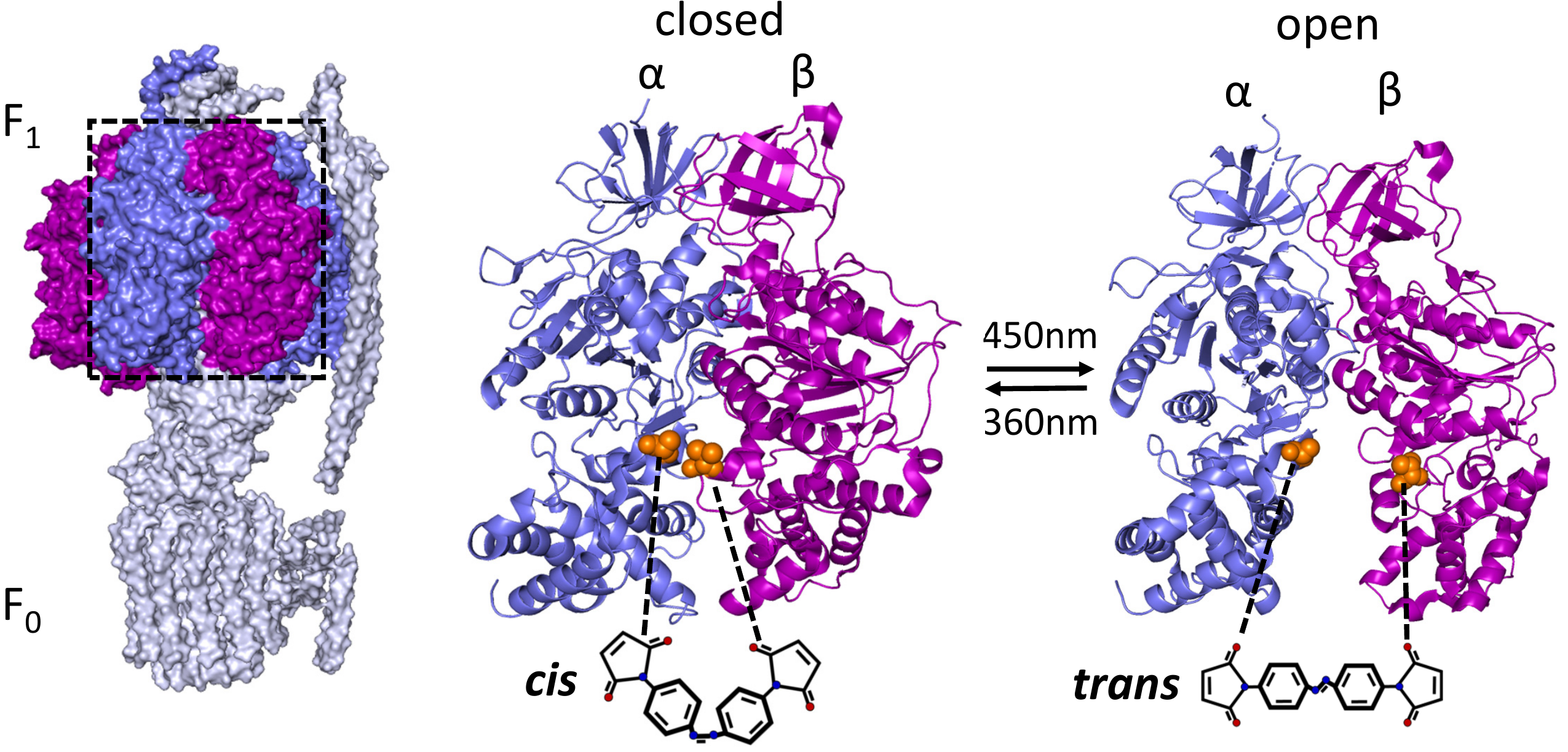}
	\caption{Photoswitchable ATPase, targeting the hinge motion of the $\alpha$ and $\beta$-subunits.\cite{Hoersch2016} The golden spheres represent the atoms of residues which are mutated to cysteines in the $\alpha$A380C/$\beta$V409C variant, which revealed the biggest difference in activity. The picture has been produced from pdb entries 5dn6 and 3oaa.}\label{figATPase}
\end{figure*}

\section{Outlook: Azobenzene Photocontrol of Biological Activity}

So far, we concentrated on structural aspects, but at the end of the day, the goal is to control biological activity. Many examples of such constructs have been described in literature,\cite{Willner1991,James2001,Liu1997,Hamachi1998,Schierling2010, Hoersch2016, Blacklock2018, Yamada2007, hull2018,Borowiak2015,Ritterson2013,Beharry2011a,Zhang2010,Nevola2013,Borowiak2015}
which all could be potential candidates for time-resolved studies, as they  incorporate the necessary ultrafast switch - azobenzene. We will discuss a few of those examples more closely in the following.

Some of the earliest attempts to photocontrol  enzyme activity was conducted in fact on the RNase~S complex, albeit with a monofunctional construct as sketched in Fig.~\ref{figPhotoswitching}a. To that end, an unnatural azobenzene-bearing amino acid (phenylazophenylalanine) was incorporated into the S-peptide by means of a standard solid phase peptide synthesis at several positions, modulating the binding affinity between the S-peptide and S-protein and thereby the enzymatic activity of the RNase~S complex.\cite{James2001, Liu1997, Hamachi1998} However, the effect was modest with less than a 5 fold difference in activity, supposedly due to the modest impact of a photoswitchable side chain that does not affect the backbone structure of the peptide. Bifunctional cross-linking as in Fig.~\ref{figMolecules}d is expected to have a bigger effect on the enzymatic activity.

An illustrative example of such a bifunctional control has been demonstrated in the \textit{in vitro} study of a photoswitchable calcium-binding protein, cadherin E.\cite{Ritterson2013} In total 11 versions were synthesized, differing by the relative positions of the two cysteines to which the photoswitch was linked, and the most potent one was further characterized. A 18-fold difference in calcium binding affinity was achieved when affecting the structure of calcium binding loop. \textit{Trans}-to-\textit{cis} isomerization of the photoswitch translated directly into the dimerization propensity of cadherin. As cadherins play a direct role in cell-cell contacts, this study shows the potential for a control of cellular adhesion. Similar design strategies were applied in Ref.~\onlinecite{Schierling2010} with a motivation to control the enzymatic activity of PvuII restriction endonuclease.

In Ref.~\onlinecite{Hoersch2016}, an even larger protein construct was targeted. The ATPase molecular machinery was cross-linked with the azobenzene photoswitch between the $\alpha$ and $\beta$-subunits of the F$_1$ part of ATP synthase (see Fig.~\ref{figATPase}). The influence of light on the ATP-hydrolyzing activity was tested for 4 mutants with different positions of the cysteine pair. The highest effect was achieved when \textit{cis}-to-\textit{trans} photoswitching induces a large difference in the flexibility of the active site, enabling the opening and closing of the F$_1$ subunit.

In the \textit{in vivo} study of Ref.~\onlinecite{Borowiak2015}, the biological activity of tubulin, the protein that constitutes the microtubule cytoskeleton, was controlled indirectly by modifying its polymerization inhibitor. To that end, an analogue of the small molecule inhibitor based on an azobenzene substituted with functional methoxy groups was used, important for inhibitory activity. Only in the \textit{cis}-configuration of the azobenzene, methoxy groups were spatially arranged in an appropriate fashion to mimic the inhibitor. Upon switching to the \textit{trans}-state, the inhibitory activity was abolished. This compound was applied to the cell culture and a complete ``on/off'' switching of the microtubule assembly was observed, allowing for a control of mitosis and cell death with single-cell precision.

The numerous efforts to control biological activity by azobenzene switches could be enriched by including a time-resolution, i.e., see how the time-dependent structural changes affect the activity of the proteins. All the processes illustrated above could be followed by time-resolved methods upon ultrafast photoswitching, addressing what happens with the protein in real-time. This would be a way to connect the hierarchy of timescales with the mechanisms of e.g. enzyme catalysis, microtubular assembly, endocytosis and many other biological processes. Presumably, as the size of a protein system becomes larger even slower timescales will become relevant, but we expect that fast timescales remain equally important. Connecting time-resolved studies with biological activities would go significantly beyond our current understanding of the structure/function paradigm.

\section{Conclusion}

Understanding processes as complex as the dynamical nature of proteins requires a joint effort between several scientific fields. Introducing time-resolved  techniques in various spectral regions is necessary to extend in-equilibrium studies of protein dynamics. Besides spectroscopic techniques, as discussed here, we have seen a tremendous advance in ultrafast time-resolved X-ray scattering in recent years, and various naturally photoactive proteins have been studied thus far.\cite{schotte03,Knapp2006,Kern2013,Nogly2018,Standfuss2019,Skopintsev2020} The naturally photoactive proteins are once again laying the foundation for studying artificially controllable systems, and we expect to see first results in the years to follow.

In the meantime, photoswitches suitable for \textit{in vivo} applications, namely ones that isomerise on longer wavelengths (red to near-infrared)\cite{Dong2015,Dong2017} and are resistant to reductive environments and hydrolysis,\cite{Dong2015} are being developed. First attempts to apply azobenzene photocontrol in living systems have been made and its potential use in therapeutic targeting has been demonstrated.\cite{Beharry2011a,Zhang2010,Nevola2013,Borowiak2015}  As manipulating protein activity with azobenzene photoswitches is getting more adaptable and as its applications are broadening, optogenetics is emerging as a way of making proteins photoresponsive by genetic manipulation of naturally occurring photoresponsive proteins or domains.\cite{deisseroth2011,Hausser2014} It usually designs novel photoresponsive proteins based on, e.g. retinal (or other chromophores) binding domains and their linking to other proteins on the genetic level. Neuronal activity can be regulated in this way, as well as many other aspects of cellular function. A strategy to involve azobenzene for optogenetic applications involves the method of codon expansion.\cite{Hoppmann2014, Hoppmann2015}. It is based on an engineered pair of tRNA and tRNA synthetase, which incorporates an azobenzene-bearing unnatural amino acid during the protein translation directly in living cells.

In any case, azobenzene photocontrol currently offers the most versatile approach and offers answers to fundamental questions tackling the time propagation of biological processes. Using the ultrafast perturbation of a photocontrollable system as a trigger in combination with transient spectroscopy techniques allows following events supervening in real-time. With this novel approach, we can finally start to shed light on the aspect of time, which to a certain extent is a forgotten dimension in biomolecular interactions. We can monitor signals as they propagate from their origin throughout a system, and reveal the  dynamical nature of proteins. A complete dynamical depiction will help to unravel many undisclosed phenomena in biochemistry, where the time component must be included, e.g. propagation of allosteric signals in proteins, direct discrimination of protein-protein interaction mechanisms (conformational selection vs induced fit), and coupled binding/folding in intrinsically disordered proteins.\\

\noindent\textbf{Acknowledgement:} We thank all our coworkers in this project, in particular Brigitte Buchli, Steven Waldauer, Reto Walser, Oliver Zerbe, Rolf Pfister, Klemens Koziol, Philip J. M. Johnson, Claudio Zanobini, Jeannette Ruf, David Buhrke, and as well as the groups of Gerhard Stock, Amedeo Caflisch and Ben Schuler for their numerous contributions to this works. We also thank Andrew Woolley, who got us started in an initial phase of the project. The work has been supported by an European Research Council (ERC) Advanced Investigator Grant (DYNALLO) as well as the Swiss National Science Foundation (SNF) through the NCCR MUST and Grants 200021\_165789/1 and 200020B\_188694/1.\\

%\bibliographystyle{naturemag}
%\bibliography{../../library}

\newpage
\clearpage

\end{document}